# Elasticity and hydrodynamics of quasicrystals with 7-, 14-, 9- and 18-fold symmetries

Tian You Fan

School of Physics, Beijing Institute of Technology, P.O.Box 327, Beijing 100081, China

E-mail: tyfan@bit.edu.cn; tyfan2006@yahoo.com.cn

**Abstract** This letter reports theory of elasticity and hydrodynamics of quasicrystals with 7-, 14-, 9- and 18-fold symmetries in solid phase, in which the 6-dimensional embedding space concept is used. Based on the concept and the Landau-Anderson symmetry breaking principle and the Lubensky hydrodynamics, the governing equations for the above quasicrystal systems have been set up, the solving procedure for the initial-boundary value problems of the equations is discussed as well. Though the discussion is concerned only for the quasicrystals, in particular for one of 18-fold symmetry, in solid phase, it may be a basis for those in soft matter phase observed recently.

The quasicrystals were observed in binary and ternary metallic alloys for three-dimensional icosahedral quasicrystals [1] first and two-dimensional 5-, 10-, 8- and 12-fold symmetry quasicrystals [2] then. Recently the structures were discovered in colloidal solutions [3], polymers [4-6] and nanoparticle mixture [7]. Especially the work reported by Ref [3] has aroused the great interest of researchers, because the 18-fold quasicrystal was the first observed. Before the experimental discovery of 12- and 18-fold quasicrystals, Denton and Loewen [8] theoretically predicted the existence of quasicrystals in colloids.

From the view point of symmetry, the possible 7-，14-，9- and 18-fold quasicrystals in solid phase have been predicated by Hu et al [9] in an earlier time. Though the quasicrystals with these symmetries in solid phase have not been observed so far, the discovery of quasicrystals with 18-fold symmetry in colloid phase suggests the significance for studying 7-，14-，9- and 18-fold quasicrystals. If we explore the physical and mechanical properties of these solid quasicrystals, it will help one understanding those of soft matter quasicrystals. For this purpose we discuss the elasticity and hydrodynamics of 7-，14-，9- and 18-fold quasicrystals in solid phase in this letter.

Bak [10,11] and Lubensky et al [12-16] have done the pioneering work for elasticity of quasicrystals.

Anderson [17] introduced the Landau symmetry breaking concept [12] and proved for crystals if the density of periodic crystals can be expressed by a Fourier series (the expansion exists due to the periodicity of the structure in three-dimensional lattice or reciprocal lattice)

$$\rho(\mathbf{r}) = \sum_{\mathbf{G} \in L_R} \rho_{\mathbf{G}} \exp\{i\mathbf{G}\cdot\mathbf{r}\} = \sum_{\mathbf{G} \in L_R} |\rho_{\mathbf{G}}| \exp\{-i\Phi_{\mathbf{G}} + i\mathbf{G}\cdot\mathbf{r}\}$$

where $\mathbf{G}$ is a reciprocal vector, and $L_R$ the reciprocal lattice in three-dimensional space, $\rho_{\mathbf{G}}$ is a complex number

$$\rho_{\mathbf{G}} = |\rho_{\mathbf{G}}| e^{i\Phi_{\mathbf{G}}}$$

with an amplitude $|\rho_{\mathbf{G}}|$ and phase angle $\Phi_{\mathbf{G}}$, due to $\rho(r)$ being real, $|\rho_{\mathbf{G}}| = |\rho_{-\mathbf{G}}|$ and $\Phi_{\mathbf{G}} = -\Phi_{-\mathbf{G}}$, then the order parameter is

$$\eta = |\rho_{\mathbf{G}}|$$

Anderson pointed further out that for crystals the phase angle $\Phi_{\mathbf{G}}$ contains the phonon $\mathbf{u}$, i.e.,

$$\Phi_{\mathbf{G}} = \mathbf{G}\cdot\mathbf{u} \qquad (1)$$

in which both $\mathbf{G}$ and $\mathbf{u}$ are in three-dimensional physical space. If we consider only the phonetic branch of phonon, then here $\mathbf{u} = (u_x, u_y, u_z)$ can be understood as displacement field in macroscopic sense.

Bak[10,11] and Lubensky et al [12-16] extended the Landau-Anderson elementary excitation principle to quasicrystals, i.e., there is a set of $N$ base vectors, $\{\mathbf{G}_n\}$, so that each $\mathbf{G} \in L_R$ can be written as $\sum m_n \mathbf{G}_n$ for integers $m_n$. Furthermore $N = kd$, where $k$ is the number of the mutually incommensurate vectors in the $d$-dimensional quasicrystal. In general $k = 2$. A convenient parametrization of the phase angle is given for the present case

$$\Phi_n = \mathbf{G}_n^{\|} \cdot \mathbf{u} + \mathbf{G}_n^{\perp} \cdot \mathbf{w} \qquad (2)$$

in which $\mathbf{G}_n^{\|}$ is the reciprocal vector in the physical space $E_{\|}^3$ just mentioned and $\mathbf{G}_n^\perp$ is the conjugate vector in the perpendicular space $E_\perp^3$, $\mathbf{u}$ can be understood similar to the phonons like that in conventional crystals, mentioned above, while $\mathbf{w}$ can be understood the phason degrees of freedom in the quasicrystal. Note that, from the hydrodynamic sense, $\mathbf{w}$ presents different behaviour from that of $\mathbf{u}$, the phason vector $\mathbf{w}$ represents diffusion rather than wave propagation like that phonon vector $\mathbf{u}$ does, as pointed out by Lubensky et al [12-16]. In addition $\mathbf{u}$ and $\mathbf{w}$ belong to different irreducible representation of point group, i.e., phonons transform according to a radius vectorlike (in parallel space) representation of rotation group, whereas phasons transform according to other representation neither of which is the vectorlike. The two elementary excitations—phonon field $\mathbf{u}=(u_x,u_y,u_z)$ and phason field $\mathbf{w}=(w_x,w_y,w_z)$ form the basic field variables of elasticity of quasicrystals. This leads to two strain tensors of phonon and phason (the notations hereafter refer to Ding et al [18], Hu et at [19], Fan [20])

$$\varepsilon_{ij}=\frac{1}{2}\left(\frac{\partial u_i}{\partial x_j}+\frac{\partial u_j}{\partial x_i}\right), \quad w_{ij}=\frac{\partial w_i}{\partial x_j} \qquad (3)$$

respectively. The stress tensor corresponding to phonon strain tensor is denoted by $\sigma_{ij}$, and another one corresponding to phason strain tensor is marked by $H_{ij}$ and it follows that the generalized Hooke's law

$$\left.\begin{array}{l}\sigma_{ij}=\dfrac{\partial F}{\partial \varepsilon_{ij}}=C_{ijkl}\varepsilon_{kl}+R_{ijkl}w_{kl}\\ H_{ij}=\dfrac{\partial F}{\partial w_{ij}}=K_{ijkl}w_{kl}+R_{klij}\varepsilon_{kl}\end{array}\right\} \quad (4)$$

where function $F=F(u,w)$ is the elastic free energy (or the strain energy density) of the system, $C_{ijkl}$ the elastic constant tensor of phonon field, $K_{ijkl}$ the one of phason field, $R_{ijkl}$ the one for phonon-phason coupling ($uw$ coupling) field. This discussion holds for 5-, 10-, 8-and 12-fold two-dimensional quasicrystals and three-dimensional icosahedral quasicrystals observed in metallic alloys phases. These belong to the pioneering work of Bak, Lubensky etc, Ding et al [18,19] gave a summary on the work, Fan [20] further developed the summery of Ding et al, and gave systematic solving and a lot of applications

The 7-, 14, 9- and 18-fold quasicrystals (if they exist) belong to two-dimensional quasicrystals, but they are quite different from the 5-, 10-, 8- and 12-fold quasicrystals from view point of symmetry. To reveal the structures and symmetries of these four kinds of quasicrystals, it must use the so-called 6-dimensional embedding space concept. According to the group representation theory Hu et al [9] put forward the generalized Landau-Anderson equation, i.e., (1) should be extended as

$$\Phi_n = \mathbf{G}_n^{\parallel}\cdot\mathbf{u}+\mathbf{G}_n^{\perp 1}\cdot\mathbf{v}+\mathbf{G}_n^{\perp 2}\cdot\mathbf{w} \quad (5)$$

in which $\mathbf{G}_n^{\parallel}$ is the reciprocal vector just discussed previously, but in the physical space $E_{\parallel}^2$ of two-dimension, $\mathbf{G}_n^{\perp 1}$ is the first conjugate vector in the first perpendicular space $E_{\perp 1}^2$ of two dimension, $\mathbf{G}_n^{\perp 2}$ is the second conjugate vector in the second perpendicular space $E_{\perp 2}^2$ of two dimension. This means that the number $k$ of the mutually incommensurate vectors in the $d$-dimensional quasicrystal should be here $k=3$ in equation $N=kd$. In this case $\mathbf{u}$ can be understood similar to the phonon like that in previous section, but with two-dimension, i.e., $\mathbf{u}=(u_x,u_y)$, and the corresponding strain tensor is still defined by the first equation of equations (3). At meantime there are two vertical spaces, each of which contains a phason excitation, the first phason field $\mathbf{v}=(v_x,v_y)$, and the second phason field $\mathbf{w}=(w_x,w_y)$, respectively, the detail refer to [9]. The associated strain tensor with the first phason field is defined by

$$v_{ij}=\frac{\partial v_i}{\partial x_j} \qquad (6)$$

The strain tensor associated with the second phason field has been defined by second equation of equations (3) already.

Denote the stress tensor associated with strain tensor $v_{ij}$ to be $\tau_{ij}$ and the elastic constant tensor to be $T_{ijkl}$. Accordingly there are another phonon-phason coupling ($uv$ coupling) field with the coupling elastic constant tensor $r_{ijkl}$ and phason-phason ($vw$ coupling) field with the coupling elastic constant tensor $G_{ijkl}$. So that the generalized Hooke's law should be extended as

$$\left.\begin{aligned}\sigma_{ij} &= \frac{\partial F}{\partial \varepsilon_{ij}} = C_{ijkl}\varepsilon_{kl} + r_{ijkl}v_{kl} + R_{ijkl}w_{kl} \\ \tau_{ij} &= \frac{\partial F}{\partial v_{ij}} = T_{ijkl}v_{kl} + r_{klij}\varepsilon_{kl} + G_{ijkl}w_{kl} \\ H_{ij} &= \frac{\partial F}{\partial w_{ij}} = K_{ijkl}w_{kl} + R_{klij}\varepsilon_{kl} + G_{klij}v_{kl}\end{aligned}\right\} \quad (7)$$

in which $F = F(u,v,w)$ denotes the strain energy density of the system.

By extending the theory of Lubensky et al [12-16], Rochal and Norman [21], Fan [20] we can find the equations of motion of elasto-dynamics for the 7-, 14, 9- and 18-fold quasicrystals, i.e.,

$$\left.\begin{aligned}\rho\frac{\partial^2 u_i}{\partial t^2} &= \frac{\partial \sigma_{ij}}{\partial x_j} \\ \kappa_v \frac{\partial v_i}{\partial t} &= \frac{\partial \tau_{ij}}{\partial x_j} \\ \kappa_w \frac{\partial w_i}{\partial t} &= \frac{\partial H_{ij}}{\partial x_j}\end{aligned}\right\} \quad (8)$$

in which $\rho$ denotes the mass density, $\kappa_v = 1/\Gamma_v, \kappa_w = 1/\Gamma_w$, $\Gamma_v$ and $\Gamma_w$ are the kinetic coefficients of the first and second phason fields, respectively. The concept of kinetic coefficient was introduced by Lubensky et al [12-16], we here make an extension for phason field $v = (v_x, v_y)$. In this case, apart from elastic constants, the hydrodynamic parameters $\rho$, $\Gamma_v$ and $\Gamma_w$ are important too. In this sense, the elastodynamics of quasicrystals is connected essentially to hydrodynamics.

The equations (3), (6), (7) and (8) are governing equations of elasto-dynamics of the 7-, 14, 9- and 18-fold quasicrystals in solid phase.

The governing equations of hydrodynamics of these quasicrystals are equations (3), (6), (8) and the equation of motion as below

$$\left. \begin{array}{l} \dfrac{\partial(\rho V_i)}{\partial t} + V(\nabla \rho V_i) = \dfrac{\partial}{\partial x_j}\left(\sigma_{ij} + \sigma'_{ij}\right) \\[6pt] \dfrac{\partial \rho}{\partial t} + \dfrac{\partial(\rho V_j)}{\partial x_j} = 0 \\[6pt] \dfrac{\partial u_i}{\partial t} + V(\nabla u_i) + \Gamma_u \dfrac{\partial \sigma_{ij}}{\partial x_j} - V_i = 0 \\[6pt] \dfrac{\partial v_i}{\partial t} + V(\nabla v_i) + \Gamma_v \dfrac{\partial \tau_{ij}}{\partial x_j} = 0 \\[6pt] \dfrac{\partial w_i}{\partial t} + V(\nabla w_i) + \Gamma_w \dfrac{\partial H_{ij}}{\partial x_j} = 0 \end{array} \right\} \quad (9)$$

where $V = (V_x, V_y, V_z)$ denotes the velocity vector and

$$\sigma'_{ij} = \eta_{ijkl}\left(\dfrac{\partial V_k}{\partial x_l} + \dfrac{\partial V_l}{\partial x_k}\right) \quad (10)$$

the stress tensor due to viscosity of quasicrystals, and $\eta_{ijkl}$ the viscosity coefficient tensor. The equations (9) is an extension of hydrodynamics equations of Lubensky et al [12-16], they derived them in terms of the Poisson bracket method. We here make some simplifications, i.e., some small terms are omitted. In equations (9) the mass density $\rho$ is a function of time and spatial coordinates, in general, does not like that in equations (8). Enlarging of number of field variables and nonlinearity in hydrodynamics make the problem to be more complicated than that of elastodynamics.

One of the differences of the elasticity between individual quasicrystal systems lies in the material constants, which can be referred to Table 1 as below.

Table 1 Elastic constants for 7-, 14-, 9- and 18-fold quasicrystals

| Number of nonzero independent constants | 7-fold | 14-fold | 9-fold | 18-fold |
|---|---|---|---|---|
| $C_{ijkl}$ | 2 | 2 | 2 | 2 |
| $T_{ijkl}$ | 2 | 2 | 2 | 2 |
| $K_{ijkl}$ | 2 | 2 | 2 | 2 |
| $r_{ijkl}$ | 0 | 0 | 0 | 0 |
| $R_{ijkl}$ | 1 | 1 | 0 | 0 |
| $G_{ijkl}$ | 1 | 1 | 1 | 1 |

We here give a description for the elastic constants, for the most important 18-fold quasicrystals in solid phase as follows

$$C_{ijkl} = L\delta_{ij}\delta_{kl} + M(\delta_{jk}\delta_{jl} + \delta_{il}\delta_{jk}) \quad (i,j,k,l=1,2) \quad (11)$$

$$L = C_{12}, M = (C_{11} - C_{12})/2 = C_{66} \quad (12)$$

$$\left.\begin{array}{l} K_{1111} = K_{2222} = K_1, K_{1122} = K_{2211} = K_2 \\ K_{1221} = K_{2112} = K_3, K_{2121} = K_{1212} = K_1 + K_2 + K_3 \end{array}\right\} \quad (13)$$

and others are zero. The results can also be expressed

$$\begin{array}{l} K_{ijkl} = (K_1 - K_2 - K_3)(\delta_{ik} - \delta_{il}) + K_2\delta_{ij}\delta_{kl} + \\ K_3\delta_{il}\delta_{jk} + 2(K_2 + K_3)(\delta_{i1}\delta_{j2}\delta_{k1}\delta_{l2} + \delta_{i2}\delta_{j1}\delta_{k2}\delta_{l1}) \quad (i,j,k,l=1,2) \end{array} \quad (14)$$

The phason elastic constant tensor $T_{ijkl}$ has non-zero components

$$\begin{array}{l} T_{1111} = T_{2222} = T_{2121} = T_1 \\ T_{1122} = T_{2211} = -T_{2112} = -T_{1221} = T_2 \end{array} \quad (15)$$

and other $T_{ijkl} = 0$, and the expression of them by tensor of four rank is

$$\mathrm{T}_{ijkl} = T_1\delta_{ik}\delta_{jl} + T_2(\delta_{ij}\delta_{kl} - \delta_{il}\delta_{jk}) \quad (i,j,k,l=1,2) \quad (16)$$

And the first phonon-phason coupling constants due to decoupled property between phonon field u and phason field v

$$r_{ijkl} = 0 \qquad (17)$$

and the second phonon-phason coupling elastic constants are also vanish, i.e.,

$$R_{ijkl} = 0 \qquad (18)$$

and phason-phason coupling elastic constants

$$G_{ijkl} = G(\delta_{i1}-\delta_{i2})(\delta_{ij}\delta_{kl} - \delta_{il}\delta_{jk} + \delta_{iljk}) \qquad (i,j,k,l=1,2) \qquad (19)$$

these are originated from to Hu et al [9].

The viscosity coefficient tensor is also different from the different quasicrystal system, can be taken as a scalar quantity for simplicity.

The equations (3), (6), (7) and (8) are linear, the solving is relatively simpler, the solutions for stationary state (i.e., $\partial/\partial t = 0$) can be obtained through the Fourier analysis, similar to those for quasicrystals with 5-, 8-, 10- and 12-fold symmetries, refer to [20]. The equations (3), (6), (7) and (9) are nonlinear, the solving is difficult. If we take linearization, e.g. omitting the nonlinear terms, the numerical solutions can be obtained by finite difference method. If further consider the stationary state (i.e., $\partial/\partial t = 0$), some analytic solutions are available.

Though the 7-, 14-, 9- and 18-fild quasicrystals in solid phase are not observed up to now, the present work may help the solving for those problems in quasicrystals in soft matter. For the elasticity and hydrodynamics, solutions of quasicrystals of 12-fold and 18-fold in soft matter will be reported in the subsequent letter, in which the physical model and mathematical method are a development of those of the present work.

After writing the manuscript, the author obtained information, that the quasicrystals with 36-fold symmetry in nanomaterials are observed, refer to Physics World. Com, Sept 6, 2012, this encourages us to study the elasticity and hydrodynamics of the new quasicrystals.

**Acknowledgement** The work is supported by the National Natural Science Foundation of China through grant 11272053. The author thanks the helpful discussion of Professor Hu C Z in School of Physics of Wuhan University in China.